\documentclass[preprint,floatfix,showpacs,aps]{revtex4}

\newcommand {\bea}{\begin{eqnarray}}
\newcommand {\eea}{\end{eqnarray}}
\newcommand {\be}{\begin{equation}}
\newcommand {\ee}{\end{equation}}

\usepackage{graphicx}
\usepackage{epsfig,subfigure}
\usepackage{amsmath}
\begin{document}
\def\({\left(}
\def\){\right)}
\def\[{\left[}
\def\]{\right]}

\def\Journal#1#2#3#4{{#1} {\bf #2}, #3 (#4)}
\def\RPP{{Rep. Prog. Phys}}
\def\PRC{{Phys. Rev. C}}
\def\PRD{{Phys. Rev. D}}
\def\FP{{Foundations of Physics}}
\def\ZPA{{Z. Phys. A}}
\def\NPA{{Nucl. Phys. A}}
\def\JPG{{J. Phys. G Nucl. Part}}
\def\PRL{{Phys. Rev. Lett}}
\def\PRpt{{Phys. Rep.}}
\def\PLB{{Phys. Lett. B}}
\def\AP{{Ann. Phys (N.Y.)}}
\def\EPJA{{Eur. Phys. J. A}}
\def\NP{{Nucl. Phys}}
\def\ZP{{Z. Phys}}
\def\RMP{{Rev. Mod. Phys}}
\def\IJMPE{{Int. J. Mod. Phys. E}}
\input epsf

\title{Low Densities Instability of Relativistic Mean Field Models}

\author{A. Sulaksono and T. Mart}

\affiliation{Departemen Fisika, FMIPA, Universitas Indonesia,
Depok, 16424, Indonesia}

\begin{abstract}
The effects of the symmetry energy softening of the relativistic
mean field (RMF) models on the properties of matter with neutrino
trapping are investigated. It is found that the effects are less
significant than those in the case without neutrino trapping. The
weak dependence of the equation of state on the symmetry energy is
shown as the main reason of this finding. Using different RMF
models the dynamical instabilities of uniform matters, with and
without neutrino trapping, have been also studied. The interplay
between the dominant contribution of the variation of matter
composition and the role of effective masses of mesons and
nucleons leads to higher critical densities for matter with
neutrino trapping. Furthermore, the predicted critical density is
insensitive to the number of trapped neutrinos as well as to the
RMF model used in the investigation. It is also found that
additional nonlinear terms in the Horowitz-Piekarewicz and
Furnstahl-Serot-Tang models prevent another kind of instability,
which occurs at relatively high densities. The reason is that the
effective $\sigma$ meson mass in their models increases as a
function of the matter density.
\end{abstract}
\maketitle

\newpage
\section{Introduction}
\label{sec_intro}
Recently, the dynamical stability of the uniform ground state of
multi components systems (i.e., electrons, protons, and neutrons)
at low densities  has received considerable
attentions~\cite{Pethick,Douchin,Carr,Horowitz01,Provi1,Provi2}.
This interest is motivated by the fact that the neutron star is
expected to have a solid inner crust of nonuniform neutron-rich
matter above its liquid mantle~\cite{Carr} and the mass of its
crust depends sensitively on the density of its inner edge and on
its equation of state (EOS)~\cite{Douchin}. Meanwhile, the
critical density ($\rho_c$), a density at which the uniform liquid
becomes unstable to a small density fluctuation, can be used as a
good approximation of the edge density of the crust~\cite{Carr}.
Using the Skyrme SLy effective interactions,  Ref.~\cite{Douchin}
found the inner edge of the crust density to be  $\rho_{\rm edge}$
= 0.08 $\rm fm^{-3}$. Reference~\cite{Carr} generalized the
dynamical stability analysis of Ref.~\cite{Horo1} in order to
accommodate the various nonlinear terms in the relativistic mean
field (RMF) model of Horowitz-Piekarewicz~\cite{Horowitz01}. They
found a strong correlation between $\rho_c$ in neutron star and
the symmetry energy ($a_{\rm sym}$) which leads to a linear
relation between   $\rho_c$ and the skin thickness of finite
nuclei.

These results suggested that a measurement of the neutron radius
in $^{208}{\rm Pb}$ will provide useful information on the
$\rho_c$ \cite{Carr,Horowitz01}. Recently, a new version of the
RMF model based on effective field theory  (ERMF) has been
proposed by Furnstahl, Serot, and Tang~\cite{Furn96}. The
predictive power of the model in a wide range of densities is
quite impressive (see the review articles~\cite{sil,serot} for the
details). On the other hand, an adjustment of the isovector-vector
channel of the ERMF model in order to achieve a softer density
dependence of $a_{\rm sym}$ at high densities has been done in
Ref.~\cite{anto05}.

Similar to the previous calculation done in~\cite{Horowitz01}, but
using a different RMF model, it is found that by adjusting that
channel a lower proton fraction ($Y_p$) in high density neutron
star matter can be achieved without significantly changing the
bulk properties~\cite{anto05}. It is well known that $Y_p$ is
related to the threshold of the direct URCA cooling process  and
the trend of the density distribution of $Y_p$ is unique for each
model which is sensitive to the different forms of nonlinear terms
used. Therefore, an investigation of the EOS and the instability
at low densities by using different RMF models could be quite
interesting. Indeed, by comparing our results with the previous
ones~\cite{Carr,Horowitz01}, we can systematically study the
influence of the form and the strength of isovector-vector
nonlinear terms on the critical density.  The presence of
electrons and the influence of the electromagnetic interaction on
the unstable modes of asymmetric matter as well as the comparison
between  dynamical instability region with the thermodynamical one
have been investigated in Refs.~\cite{Provi1,Provi2} by using the
standard RMF model within the Vlasov formalism. They observed the
important role of the Coulomb field in the large structure
formation and also the role of the electron dynamics in restoring
the large-wavelength instabilities \cite{Provi2}.

In the actual situation, muons may also be present in the stellar
matters. Their existence yields an additional electromagnetic
contribution besides the electrons and protons ones. The
significance of their contribution depends on the matter
composition. Including muons in the dynamical instability analysis
of uniform matter increases the electromagnetic effect on the
$\rho_c$ in medium. Moreover, in supernovae or protoneutron stars
neutrinos can be trapped inside, if their mean free paths are
smaller than the star radius. In general, the presence of trapped
neutrinos in matter affects the stiffness of EOS and drastically
changes the composition of neutral
matter~\cite{Prakash,Vidana,Guo,chiapparini,Bednarek}. It was also
found that neutrino trapping causes the strange
baryon~\cite{Prakash,Vidana} and kaon condensate \cite{Prakash,
Guo} to appear at higher densities as compared to the case without
neutrino trapping. In Ref.~\cite{Bednarek}, the EOS of the
strangeness rich protoneutron star and the EOS of neutron star
matter with temperature different from  zero have been already
studied by means of the ERMF model. Meanwhile, like neutron stars,
protoneutron stars have dense cores and outer
layers~\cite{Bednarek} and in supernovae inhomogeneity can appear
below the saturation density of nuclear matter
($\rho_0$)~\cite{shen,wata}. Therefore, an investigation of the
boundary between the two phases, which leads to the determination
of  $\rho_c$, becomes crucial for a realistic and complete
description of stellar matters.

In the present work, we shall extend the analysis of dynamical stability in uniform matter of
Ref.~\cite{Carr} in two ways, i.e., first, we consider the muon contribution and, second,
we shall consider all possible mixings of vector and isovector contributions caused by
the nonlinear terms in the ERMF model. The parameter set NL3 of the standard RMF~\cite{lala} model as well as the parameter sets   Z271 and  Z271* of the Horowitz-Piekarewicz one~\cite{Horowitz01} will be presented  for comparison. The results are used to study the effect of the neutrino presence on the $\rho_c$ values. Here, we only use nucleons in the baryonic sector along with $\sigma$, $\omega$, and $\rho$ mesons in the mesonic sector because $\rho_c$ usually appears at a density lower than $\rho_0$ and the strange particles appear at a density higher than $2\rho_0$. Furthermore,  in the neutrino trapping case they appear at higher density than in the case  without neutrino trapping. Therefore, we can assume the strange particles to have only minor effects on the $\rho_c$ and, as a consequence, their contributions may be neglected in this work.  On the other hand, although in the real situation the temperature of protoneutron stars is not equal to zero and the supernovae inner core can have a temperature around $T \sim$ (10-50) MeV, the zero temperature approximation can be assumed here since the temperature effect on the EOS of supernovae matter and on the maximum mass of protoneutron star~\cite{Guo,chiapparini} is smaller than that without neutrino trapping. In this approximation, the following constraints can be  used to calculate the fraction of every constituent in matter:
\begin{itemize}
\item balance equation for the chemical potentials
\be
\mu_{n}+\mu_{\nu_e} = \mu_{p}+\mu_{e},
\ee
\item conservation of the charge neutrality
\be
\rho_e +\rho_{\mu}=\rho_p,
\ee
\end{itemize}
where the total density of baryon is given by
\be
\rho_B =\rho_n+\rho_p,
\ee
while the fixed electronic-leptonic fraction $Y_{l_e}$ is defined as
\be
Y_{l_e}=\frac{\rho_e+\rho_{\nu_e}}{\rho_B}\equiv Y_e+Y_{\nu_e},
\ee
where $Y_{\nu_e}$ and  $Y_e$ are neutrino electron and electron fractions, respectively. In addition, we will revisit the EOS of matter with neutrino trapping because we want to study the effect of different adjustments in the isovector-vector sector \cite{anto05} on the EOS of matter with neutrino trapping. They were not fully explored in our previous works~\cite{anto05,anto06}.

This paper is organized as follows. In Sec. II, a brief review of the model is given. Calculation of the dielectric function is presented in Sec. III, while numerical results along with the corresponding discussions are given in Sec. IV. Finally, we give the conclusion in Sec. V.

\section{Relativistic Mean Field Models}
\label{sec_models}
Our starting point to describe the non-strange dense stellar
matter is an effective Lagrangian density for interacting
nucleons, $\sigma$, $\omega$, $\rho$ mesons as well as
noninteracting leptons. The Lagrangian density for  standard RMF
models can be found in Refs.~\cite{pg,ring,serot}. We note that in
order to modify the density dependence of $a_{\rm sym}$,
Horowitz-Piekarewicz have recently added  isovector-vector
nonlinear terms (${\mathcal L}_{\rm HP}$) in a standard  RMF
model~\cite{Horowitz01}. In the case of ERMF model, the
corresponding Lagrangian density can be found in
Refs.~\cite{Furn96, Wang00,anto05}. For the  convenience of the
reader, we write the Lagrangian density in a compact form as

\be
{\mathcal L} = {\mathcal L}_N + {\mathcal L}_M + {\mathcal L}_{\rm HP}+{\mathcal L}_{L},
\label{eq:nuclag}
\ee
where the nucleon part, up to order $\nu$ = 3, reads
\bea
{\mathcal L}_N &=&\bar{\psi}[i \gamma^{\mu}(\partial_{\mu}+i \bar{\nu}_{\mu}+i g_{\rho}  \bar{b}_{\mu}+i g_{\omega} V_{\mu})+g_A  \gamma^{\mu} \gamma^{5} \bar{a}_{\mu}\nonumber\\&-&M+g_{\sigma} \sigma]\psi-\frac{f_{\rho} g_{\rho}}{4 M }\bar{\psi}\bar{b}_{\mu \nu} \sigma^{\mu \nu}\psi,
\eea
with
\be
\psi=\left( {p \atop n}\right),  ~ ~ ~ ~\bar{\nu}_{\mu}=-\frac{i}{2}(\bar{\xi}^{\dagger}\partial_{\mu}\bar{\xi}+\bar{\xi}\partial_{\mu}\bar{\xi}^{\dagger})= \bar{\nu}_{\mu}^{\dagger},
\ee
\be
\bar{a}_{\mu}=-\frac{i}{2}(\bar{\xi}^{\dagger}\partial_{\mu}\bar{\xi}-\bar{\xi}\partial_{\mu}\bar{\xi}^{\dagger})= \bar{a}_{\mu}^{\dagger},
\ee
\be
\bar{\xi}= {\rm exp}(i \bar{\pi}(x)/f_{\pi}), ~ ~ ~ ~ \bar{\pi}(x)=\frac{1}{2} \vec{\tau}\cdot \vec{\pi}(x),
\ee
\be
\bar{\pi}(x)=\frac{1}{2} \vec{\tau}\cdot \vec{\pi}(x),
\ee
\be
\bar{b}_{\mu \nu} = D_{\mu}\bar{b}_{\nu}-D_{\nu}\bar{b}_{\mu}+i g_{\rho}[\bar{b}_{\mu},\bar{b}_{\nu}],  ~ ~ ~ ~  D_{\mu}=\partial_{\mu}+i\bar{\nu}_{\mu},
\ee
\be
V_{\mu \nu}=\partial_{\mu}V_{\nu}-\partial_{\nu}V_{\mu},
\ee
\be
\sigma^{\mu \nu}=\frac{1}{2}[\gamma^{\mu},\gamma^{\nu}].
\ee
where, $p$ and $n$ are the proton and  neutron fields, $M$  is the nucleon mass,  while  $\sigma$, $\vec{\pi}$, $V^{\mu}$, and $\vec{b}^{\mu}$ are the $\sigma$,  $\pi$, $\omega$ and $\rho$ meson fields, respectively.
The meson contribution, up to order $\nu=4$ is
\bea
{\mathcal L}_M &=&\frac{1}{4}f_{\pi}^2 {\rm Tr} (\partial_{\mu}\bar{U}\partial^{\mu}\bar{U}^{\dagger})+\frac{1}{4}f_{\pi}^2 {\rm Tr}(\bar{U} \bar{U}^{\dagger}-2)+\frac{1}{2}\partial_{\mu}\sigma\partial^{\mu}\sigma\nonumber\\ &-&\frac{1}{2} {\rm Tr}(\bar{b}_{\mu \nu}\bar{b}^{\mu \nu})-\frac{1}{4}V_{\mu \nu}V^{\mu \nu}-g_{\rho \pi \pi}\frac{2 f_{\pi}^2}{m_{\rho}^2} {\rm Tr}(\bar{b}_{\mu \nu}\bar{\nu}^{\mu \nu})\nonumber\\
&+&\frac{1}{2}\left(1+\eta_1 \frac{g_{\sigma} \sigma}{M}+\frac{\eta_2}{2} \frac{g_{\sigma}^2 \sigma^2}{M^2}\right)m_{\omega}^2 V_{\mu}V^{\mu}+\frac{1}{4!}\zeta_0 g_{\omega}^2 (V_{\mu}V^{\mu})^2\nonumber\\&+&\left(1+\eta_{\rho} \frac{g_{\sigma} \sigma}{M}\right)m_{\rho}^2 {\rm Tr}(\bar{b}_{\mu}\bar{b}^{\mu})-m_{\sigma}^2\sigma^2\left(1+\frac{\kappa_3}{3 !} \frac{g_{\sigma} \sigma}{M}+\frac{\kappa_4}{4 !} \frac{g_{\sigma}^2 \sigma^2}{M^2}\right),
\eea
where
\be
\bar{U}=\bar{\xi}^2,  ~ ~ ~ ~\bar{\nu}_{\mu \nu} = \partial_{\mu}\bar{\nu}_{\nu}-\partial_{\nu}\bar{\nu}_{\mu}+i [\bar{\nu}_{\mu},\bar{\nu}_{\nu}]=-i[\bar{a}_{\mu},\bar{a}_{\nu}].
\ee
In the mean field approximation, $\pi$ meson does not contribute.  In the Horowitz-Piekarewicz model \cite{Horowitz01} the isovector-vector nonlinear term  reads
\be
{\mathcal L}_{\rm HP}= 4 \Lambda_V  g_{\rho}^2 g_{\omega}^2 ~\vec{b}^{\mu}\cdot \vec{b}_{\mu}~ V^{\mu}  V_{\mu}.
\ee

We note that changing the $ g_{\rho}$ and $\Lambda_V$ parameters
in the Horowitz-Piekarewicz model~\cite{Horowitz01}, or $g_{\rho}$
and  $\eta_{\rho}$ parameters in the ERMF model~\cite{anto05},
affects the density dependent part of the $a_{\rm sym}$. A more
detailed procedure to adjust the density dependence of the nuclear
matter symmetry energy in RMF models can be found in
Refs.~\cite{Horowitz01,Shen05,anto05}.

For the leptons, the free Lagrangian density

\be
{\mathcal L}_{L}=\sum_{l=e^-,~\mu^-,~\nu_e,~~\nu_\mu} \bar{l}( \gamma^{\mu}\partial_{\mu}-m_l)l,
\ee
is used. In the present work, the following parameter sets  are chosen:
\begin{itemize}
\item Standard RMF model: NL3~\cite{lala},
\item Standard RMF  model plus isovector-vector nonlinear term: Z271 and Z271*~\cite{Horowitz01},
\item ERMF models: G2~\cite{Furn96} and G2*~\cite{anto05}.
\end{itemize}
The coupling constants for all parameter sets are shown in
Table~\ref{tab:params1}.

\begin{table}
\centering
\caption {Numerical values of the coupling constants used in the parameter sets.}\label{tab:params1}
  \begin{ruledtabular}
\begin{tabular}{crcrcr}
Parameter &G2~~ & NL3~~& G2*~~& Z271~~& Z271*\\\hline
$m_{\sigma}/M$        &0.554~~& 0.541~~& 0.554~~&0.495~~&0.495 \\
$g_{\sigma}/(4 \pi)$ &0.835~~& 0.813~~& 0.835~~&0.560~~&0.560 \\
$g_{\omega}/(4 \pi)$ &1.016~~& 1.024~~& 1.016~~&0.670~~&0.670 \\
$g_{\rho}/(4 \pi)$ &0.755~~& 0.712~~& 0.938~~&0.792~~&0.916 \\
$\kappa_3$     &3.247~~& 1.465~~& 3.247~~&1.325~~&1.325 \\
$\kappa_4$     &0.632~~&$-5.668$~~~~&0.632~~&31.522~~&31.522 \\
$\zeta_0$      &2.642~~& 0~~& 2.642~~&4.241~~&4.241 \\
$\eta_1$       &0.650~~& 0~~& 0.650~~& 0~~&0\\
$\eta_2$       &0.110~~& 0~~& 0.110~~& 0~~&0\\
$\eta_{\rho}$  &0.390~~& 0~~& 4.490~~& 0~~&0\\
$\Lambda_V$ &0~~& 0~~& 0~~& 0~~&0.03\\ 
\end{tabular}\\
  \end{ruledtabular}
\end{table}

\section{Dynamical Instability of Uniform Dense Stellar Matters}
\label{sec_inmat}
In order to accommodate muons and various nonlinear terms in the
ERMF  model, we have to extend the stability analyses of the
uniform ground state of Refs.~\cite{Horo1,Horo2, Carr}. To that
end, the longitudinal polarization matrix in these analyses is
modified as follows
 \bea \Pi_L = \left( \begin{array} {ccccc}
\Pi_{00}^e & 0 & 0 & 0&0\\
0& \Pi_{00}^{\mu}  & 0 & 0&0\\
0 &0& \Pi_{s}& \Pi_{m}^p &\Pi_{m}^n\\
0 &0& \Pi_{m}^p & \Pi_{00}^p & 0\\
0 &0& \Pi_{m}^n & 0 & \Pi_{00}^n
\end{array}\right),
\label{eq:longpol}
\eea
where  the scalar polarization $\Pi_{s}$ = $\Pi_{s}^p$ + $\Pi_{s}^n$ . In Eq.~(\ref{eq:longpol}), the  polarization due to the  mixing between scalar and vector terms through sigma meson is indicated by $\Pi_{m} $. In the limit of $q_0 \rightarrow 0$, the individual polarization components are \cite{Carr}
\bea
 \Pi_{s}(q,0)&=& \frac{1}{2 \pi^2} \bigg{\{} k_F E_F - \left(3 M^{*~2} + \frac{q^2}{2}\right)  ~{\rm ln} \frac{k_F+E_F}{M^*}\nonumber\\
&+&  \frac{2 E_F E^2}{q} {\rm ln} ~\Big| \frac{2 k_F - q}{2 k_F + q} \Big| -\frac{2 E^3}{q}  {\rm ln} ~\Big| \frac{q E_F - 2 k_F E}{q E_F + 2 k_F E} \Big| \bigg{\}},
\eea
for the scalar polarization,
\bea
\Pi_{m}(q,0)= \frac{M^*}{2 \pi^2}  \bigg{\{} k_F - \left(\frac{ k_F^2}{q}-\frac{q}{4}\right) {\rm ln} ~\Big| \frac{2 k_F - q}{2 k_F + q} \Big|\bigg{\}},
\eea
for the mixed scalar-vector polarization, and
\bea
\Pi_{00}(q,0)&=&-\frac{1}{ \pi^2} \bigg{\{}\frac{2}{3}  k_F E_F - \frac{q^2}{6} ~{\rm ln} \frac{k_F+E_F}{M^*}
 - \frac{E_F }{3 q} \left( M^{*~2} + k_F^2-\frac{3 q^2}{4}\right)  {\rm ln} ~\Big| \frac{2 k_F - q}{2 k_F + q} \Big|\nonumber\\ &+&\frac{E}{3 q}\left(M^{*~2} - \frac{q^2}{2}\right) {\rm ln} ~\Big| \frac{q E_F - 2 k_F E}{q E_F + 2 k_F E} \Big| \bigg{\}},
\eea
for the longitudinal polarization,  with Fermi momentum $k_F$, nucleon effective mass $M^* = M - g_{\sigma} \sigma$, Fermi energy $E_F=(k_F^2+M^{* ~2})^{1/2}$ and $E =(q^2/4+M^{* ~2})^{1/2}$. For electron ($e$) or muon ($\mu$), $M^*$ is equal to electron or muon mass, respectively.

The longitudinal meson propagator now reads
\bea
D_L = \left( \begin{array} {ccccc}
d_g & d_g& 0 & -d_g & 0\\
d_g & d_g& 0 & -d_g & 0\\
0&0& -d_s & d_{sv\rho}^+ & d_{sv\rho}^-\\
-d_g&-d_g &  d_{sv\rho}^+ & d_{33} & d_{v\rho}^-\\
0 &0& d_{sv\rho}^- &  d_{v\rho}^- & d_{44}
\end{array}\right),
\label{eq:longprop}
\eea
where $d_{sv\rho}^+ = - (d_{sv} +   d_{s\rho}$), $d_{sv\rho}^- =  -(d_{sv} -   d_{s\rho}$), $d_{v\rho}^- =  d_{v} -   d_{\rho}$, $d_{33} = d_g +  d_{v} + d_{\rho} + 2  d_{v\rho}$ and $d_{44} = d_{v} + d_{\rho} - 2  d_{v\rho}$. In this form, mixing propagators between isoscalar-scalar and isoscalar-vector ($d_{sv}$), isoscalar-vector and isovector-vector ($d_{v\rho}$), isoscalar-scalar and isovector-vector ($d_{s\rho}$) are present due to the nonlinear terms in the model, in addition to the standard  photon, omega, sigma and rho propagators ($d_g$,  $d_{v}$, $d_{s}$ and  $d_{\rho}$). These propagators are determined from the quadratic fluctuations around the static solutions which are generated by the second derivatives of energy density (${\partial^2 \epsilon}/{\partial \phi_i \partial\phi_j}$), where $\phi_i$ and $\phi_j$ are the  involved meson fields.  The energy density $\epsilon$ derived from Eq.~(\ref{eq:nuclag}) reads
\bea
\epsilon&=&\frac{2}{(2 \pi)^3} \sum_{i=p,n,e,\mu,\nu_e,\nu_\mu} \int d^3 k_i E_i (k_i)+g_\omega V_0 (\rho_p+\rho_n)+\frac{1}{2}g_\rho b_0 (\rho_p-\rho_n)\nonumber\\
&-&\frac{1}{4} c_3  V_0^4 - \frac{1}{2} m_\omega^2 V_0^2 - d_2 \sigma V_0^2 - \frac{1}{2} d_3  \sigma^2 V_0^2\nonumber\\&+&\frac{1}{2} m_\sigma^2 \sigma^2+\frac{1}{3} b_2\sigma^3+ \frac{1}{4} b_3\sigma^4\nonumber\\ &-&\frac{1}{2} m_\rho^2 b_0^2 - f_2 \sigma b_0^2 -  \frac{1}{2}\tilde{\Lambda}_s  b_0^2 \sigma^2 -  \frac{1}{2} \tilde{\Lambda}_v  b_0^2 V_0^2.
\label{eq:edens}
\eea

From Eq.~(\ref{eq:edens}) we can obtain the explicit form of all
contributions to the longitudinal propagator. Note that the energy
density of the ERMF model can be obtained from
Eq.~(\ref{eq:edens}) by using the following explicit expressions
of the coupling constants: 
\bea b_2&=&\frac{g_\sigma \kappa_3
m_\sigma^2}{2 M}, ~ ~ ~ b_3=\frac{g_\sigma^2 \kappa_4
m_\sigma^2}{6 M^2},  ~ ~ ~
f_2=\frac{g_\sigma \eta_\rho m_\rho^2}{2 M},\nonumber\\
d_2&=&\frac{g_\sigma \eta_1 m_\omega^2}{2 M}, ~ ~ ~
d_3=\frac{g_\sigma^2 \eta_2 m_\omega^2}{2 M^2},  ~ ~ ~
c_3=\frac{g_\omega^2 \xi_0}{6 },\nonumber\\\tilde{\Lambda}_s &=&
\tilde{\Lambda}_v=0. \eea

On the other hand, if we set the  coupling constants in Eq.~(\ref{eq:edens}) to
\be
f_2 = d_2 = d_3 = 0, ~ ~ ~ \tilde{\Lambda}_s = 2 \Lambda_s g_\rho^2 g_\sigma^2, ~~ ~ \tilde{\Lambda}_v = 2 \Lambda_v g_\rho^2 g_\omega^2,
\ee
we will obtain the energy density of  the Horowitz-Piekarewicz model~\cite{Horowitz01}. The mesons effective masses and the mixing polarizations calculated from the energy density [Eq.~(\ref{eq:edens})] are given by
\bea
m_\sigma^{*~2}&=& \frac{\partial^2 \epsilon}{\partial^2 \sigma}=m_\sigma^2+2 b_2 \sigma+3 b_3 \sigma^2 - d_3  V_0^2 -  \tilde{\Lambda}_s  b_0^2,\nonumber\\ m_\omega^{*~2}&=& -\frac{\partial^2 \epsilon}{\partial^2 V_0}=m_\omega^2+2 d_2\sigma+d_3 \sigma^2 + 3 c_3  V_0^2 +  \tilde{\Lambda}_v  b_0^2, \nonumber\\m_\rho^{*~2}&=& -\frac{\partial^2 \epsilon}{\partial^2 b_0}=m_\rho^2+2 f_2\sigma + \tilde{\Lambda}_s \sigma^2 +  \tilde{\Lambda}_v V_0^2,\nonumber\\
\label{eq:meseffmass}
\eea
and
\bea
\Pi_{\sigma \omega}^0&=&-\frac{\partial^2 \epsilon}{\partial \sigma \partial V_0}= 2 d_2 V_0 + 2 d_3 \sigma V_0,\nonumber\\ \Pi_{\sigma \rho}^0&=&-\frac{\partial^2 \epsilon}{\partial \sigma \partial b_0}= 2 f_2 b_0 + 2 \tilde{\Lambda}_s \sigma b_0,\nonumber\\ \Pi_{\omega \rho}^{0 0}&=&\frac{\partial^2 \epsilon}{\partial V_0 \partial b_0}= -2 \tilde{\Lambda}_v V_0 b_0.\nonumber\\
\label{eq:polar}
\eea
By substituting Eqs.~(\ref{eq:meseffmass}) and~(\ref{eq:polar}) in the $\sigma$, $\omega$, and $\rho$ propagators,
\bea
d_s=\frac{g_{\sigma}^2 (q^2+m_\omega^{*~2})(q^2+m_\rho^{*~2})}{(q^2+m_\omega^{*~2})(q^2+m_\rho^{*~2})(q^2+m_\sigma^{*~2})+(\Pi_{\sigma \omega}^0)^2(q^2+m_\rho^{*~2})+(\Pi_{\sigma \rho}^0)^2 (q^2+m_\omega^{*~2})},
\label{eq:d1}
\eea

\bea
d_v=\frac{g_{\omega}^2 (q^2+m_\sigma^{*~2})(q^2+m_\rho^{*~2})}{(q^2+m_\omega^{*~2})(q^2+m_\rho^{*~2})(q^2+m_\sigma^{*~2})+(\Pi_{\sigma \omega}^0)^2(q^2+m_\rho^{*~2})-(\Pi_{\omega \rho}^{00})^2 (q^2+m_\sigma^{*~2})},
\label{eq:d2}
\eea

\bea
d_\rho=\frac{1/4 g_\rho^2 (q^2+m_\sigma^{*~2})(q^2+m_\omega^{*~2})}{(q^2+m_\omega^{*~2})(q^2+m_\rho^{*~2})(q^2+m_\sigma^{*~2})+(\Pi_{\sigma \rho}^0)^2(q^2+m_\omega^{*~2})-(\Pi_{\omega \rho}^{00})^2 (q^2+m_\sigma^{*~2})},
\label{eq:d3}
\eea
and in the mixing propagators,
\bea
d_{sv}=\frac{g_{\sigma} g_{\omega} \Pi_{\omega \sigma}^{0}(q^2+m_\rho^{*~2})}{H(q,q_0=0)},
\label{eq:d4}
\eea

\bea
d_{s\rho}=\frac{1/2 g_\rho g_{\sigma} \Pi_{\sigma \rho}^{0}(q^2+m_\omega^{*~2})}{H(q,q_0=0)},
\label{eq:d5}
\eea

\bea
d_{v \rho}&=&\frac{1/2 g_\rho g_{\omega} \Pi_{\omega \rho}^{00}(q^2+m_\sigma^{*~2})}{H(q,q_0=0)},
\label{eq:d6}
\eea
with
\bea
H(q,q_0=0)&=&(q^2+m_\omega^{*~2})(q^2+m_\rho^{*~2})(q^2+m_\sigma^{*~2})+(\Pi_{\sigma \omega}^0)^2(q^2+m_\rho^{*~2})\nonumber\\&+&(\Pi_{\sigma \rho}^0)^2 (q^2+m_\omega^{*~2})-(\Pi_{\omega \rho}^{00})^2 (q^2+m_\sigma^{*~2}),
\eea
and photon's propagator
\bea
 d_g= \frac{e^2}{q^2},
\eea
and using the explicit form of each component, the longitudinal meson propagator can be obtained.

The explicit derivation of each propagator is given in Appendix~\ref{sec_mp}. Note that by setting all coupling constants in Eqs.~(\ref{eq:d1})~-~(\ref{eq:d6}) which are not required by  the Horowitz-Piekarewicz model~\cite{Horowitz01} to zero, Eqs. (8), (13) - (14) of Ref.~\cite{Carr} can be obtained.

The uniform ground state system becomes unstable to small-amplitude density fluctuations with momentum transfer $q$ when the following condition is satisfied~\cite{Carr}

\be {\rm{det}} \[1-D_L(q) \Pi_L(q,q_0=0)\] \le 0. \label{eq:det}
\ee The explicit form of Eq.~(\ref{eq:det}) is given in
Appendix~\ref{sec_det}. In the case that the density is smaller
than $\rho_0$, the critical density $\rho_c$ is the largest
density for which Eq.~(\ref{eq:det}) has a solution. In the case
that the density is larger than $\rho_0$, if any, $\rho_c$ is the
smallest density.

\section{Numerical Results and Discussions}
\label{sec_results}
\begin{figure*}
\centering
 \mbox{\epsfig{file=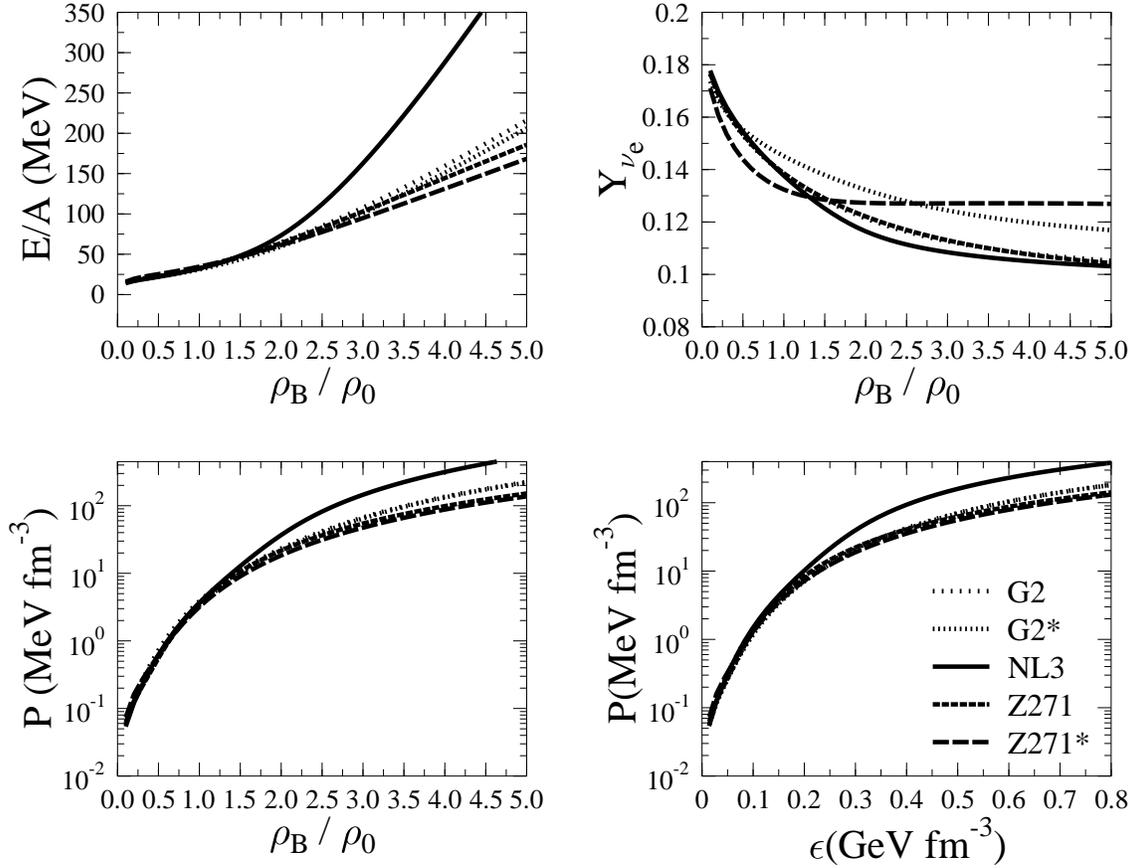,height=12.0cm}}
\caption{Binding energy per nucleon (upper left panel) and the neutrino electron fraction (upper right panel) as a function of the baryon density, as well as the equation of states (EOS) as functions of baryon density (lower left panel) and energy density (lower right panel) according to RMF models. The curves are obtained by using $Y_{l_e}=0.3$. \label{eos1}}
\end{figure*}

\begin{figure*}
\centering
 \mbox{\epsfig{file=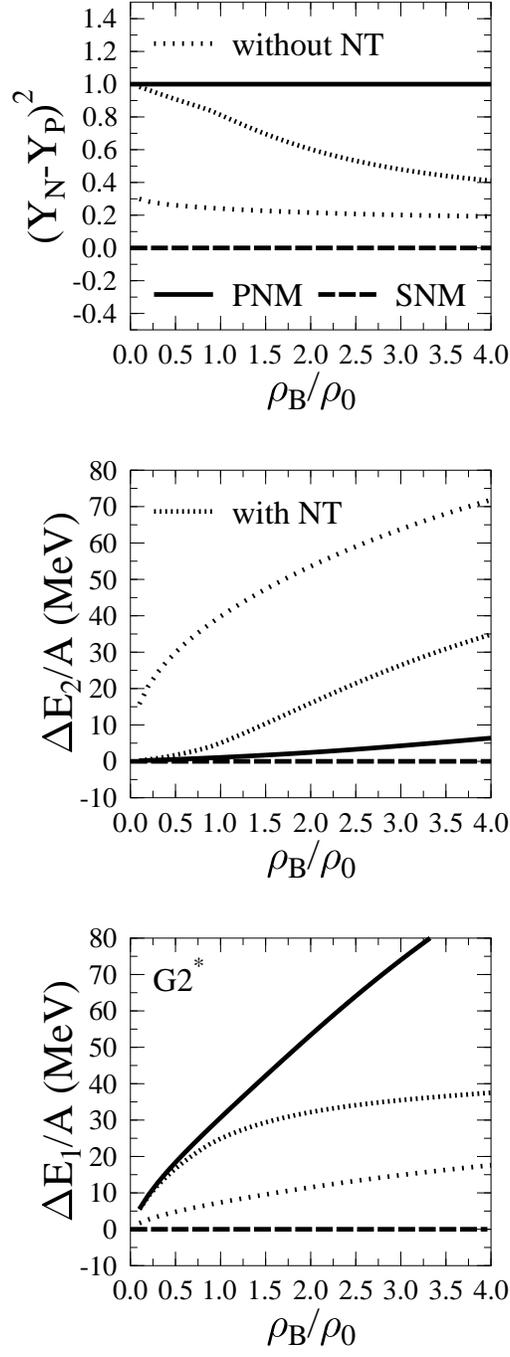,height=19.0cm}}
\caption{$\Delta E_2/A$, $\Delta E_1/A$ and $\alpha^2$ as a function of the ratio between baryon and nuclear saturation densities for matters with  and without neutrino trapping (NT). The G2* parameter set and $Y_{l_e}=0.3$ are used in obtaining these results.}
\label{tab:eden1}
\end{figure*}

Before discussing the dynamical instability of non-strange dense stellar matter,
we will discuss the effects of different  treatments in the isovector-vector sector~\cite{anto05}
 on the EOS, binding energies ($E/A$) and the neutrino electron fraction ($Y_{\nu_e}$) of matter
 with neutrino trapping. The results are shown in Fig.~\ref{eos1}. In contrast to the case
 of matter without neutrino trapping \cite{anto05}, significant differences
 in the trends of
 $E/A$ and EOS start to appear at $\rho_B/\rho_0 \approx 2.0$. Except at sufficiently high densities,
 the difference between G2 and G2*, or Z271 and Z271*,
 does not significantly show up in these trends. These results can be understood from the asymmetry expansion of
 the binding energy of the corresponding matter in the vicinity of the symmetric nuclear
 matter (SNM). The latter is given by~\cite{Shen05,Dieper,Steiner}
\bea E/A (\rho_B, \alpha)= {(E/A)}_{\rm SNM}(\rho_B, \alpha)
+\underbrace{\alpha^2 a_{\rm sym}(\rho_B)}_{\Delta E_1/A(\rho_B,
\alpha)}+ \underbrace{O(\alpha^4)}_{\Delta E_2/A(\rho_B,
\alpha)}+\Delta E_L/A(\rho_B) , \eea where $\alpha= Y_N-Y_P$ and
$\Delta E_L/A$ is the electrons and muons contribution to the
binding energy. We have found that for all parameter sets and
matters used the value of $\Delta E_L/A$ is smaller than the other
terms. The $\Delta E_2/A$ term is given by~\cite{Steiner} \bea
\Delta E_2/A (\rho_B, \alpha)=\alpha^4 Q (\rho_B)+.. . \eea The
origin and connection of the quartic term [$Q (\rho_B)$] to direct
URCA are discussed in Ref.~\cite{Steiner}. For the case of pure
neutron matter ($\Delta E_1/A = a_{\rm sym}$) or other fixed
$\alpha$ cases  it is known that  $\Delta E_2 \ll  \Delta E_1$.
This means that the convergence of the expansion is considerably
fast for these asymmetric matters but for matter with and without
neutrino trapping the situation is quite different (see Fig.
\ref{tab:eden1}). We can see that by imposing the neutrality and
$\beta$ stability conditions on the matter with and without
neutrino trapping, the corresponding asymmetry ($\alpha$) becomes
density dependent and, evidently, their $\Delta E_2/A$ are
substantially larger compared to that of the PNM, for example.
This indicates that in these cases, the convergence is
significantly slow or even can not be reached at all. Additional
constraint in the form of fixed electronic lepton fraction
($Y_{l_e}$) for the case of matter with neutrino trapping causes
the decrease of $\alpha^2$ and the convergence of binding energy
expansion are slower than those in neutrinoless matter.

\begin{figure*}
\centering
 \mbox{\epsfig{file=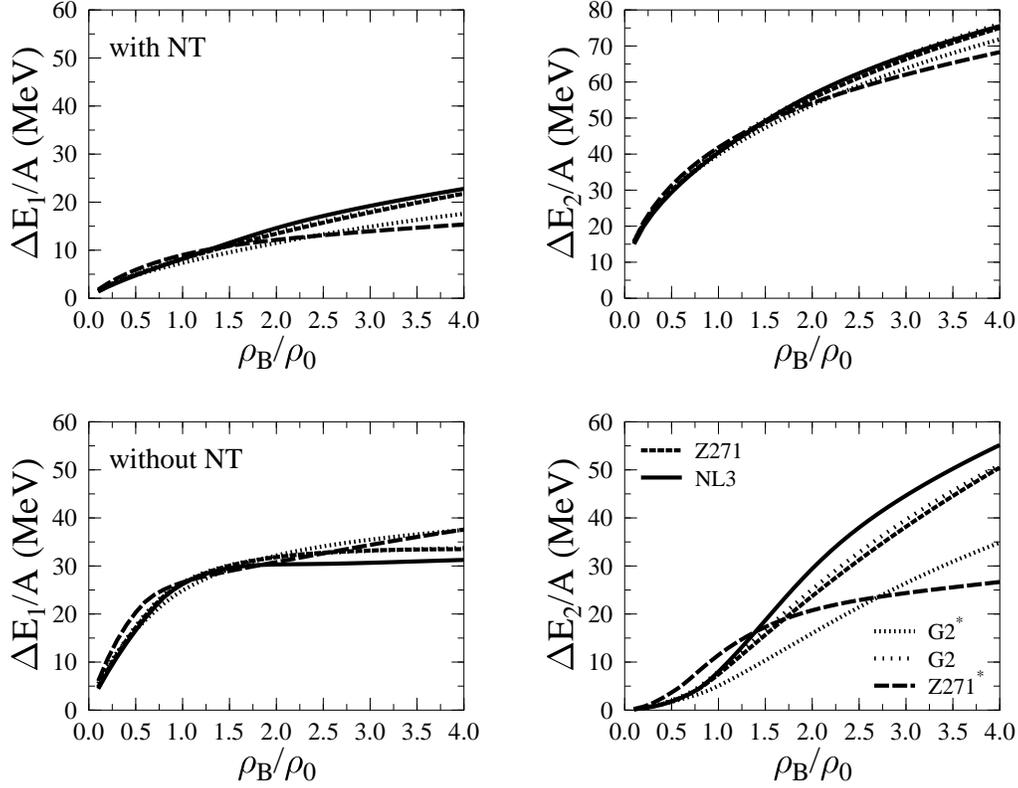,height=11.0cm}}
\caption{$\Delta E_2/A$ and $\Delta E_1/A$ as a function of the ratio between baryon and nuclear saturation densities for matters with  and without neutrino trapping (NT), obtained by using G2, G2*, Z271, Z271* and NL3 parameter sets.  All curves are obtained with $Y_{l_e}=0.3$. }
\label{tab:eden2}
\end{figure*}
\begin{figure*}
\centering
 \mbox{\epsfig{file=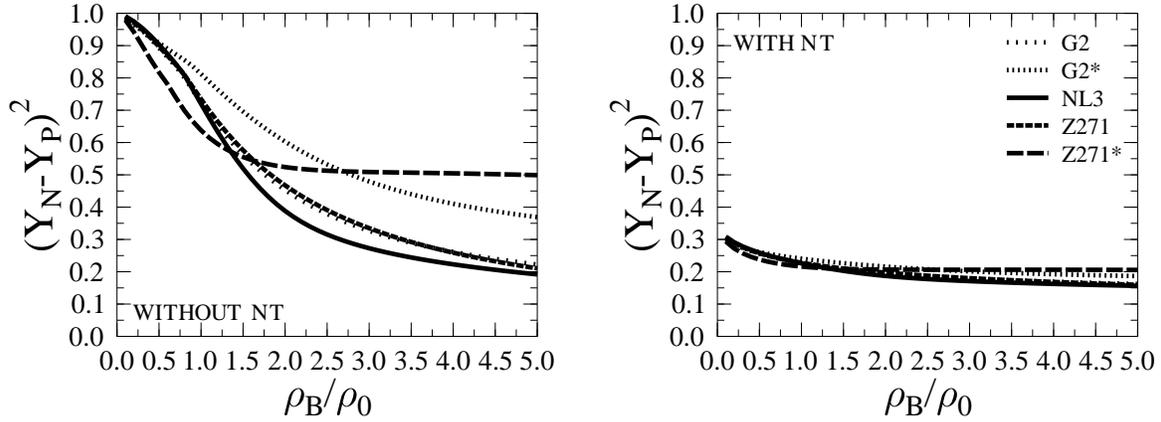,height=6.5cm}}
\caption{$\alpha^2$ as a function of the ratio between baryon and nuclear saturation densities for matters with  and without neutrino trapping (NT). The results are obtained by using G2, G2*, Z271, Z271*, NL3 parameter sets and $Y_{l_e}=0.3$.}
\label{tab:eden3}
\end{figure*}
\begin{figure*}
\centering
 \mbox{\epsfig{file=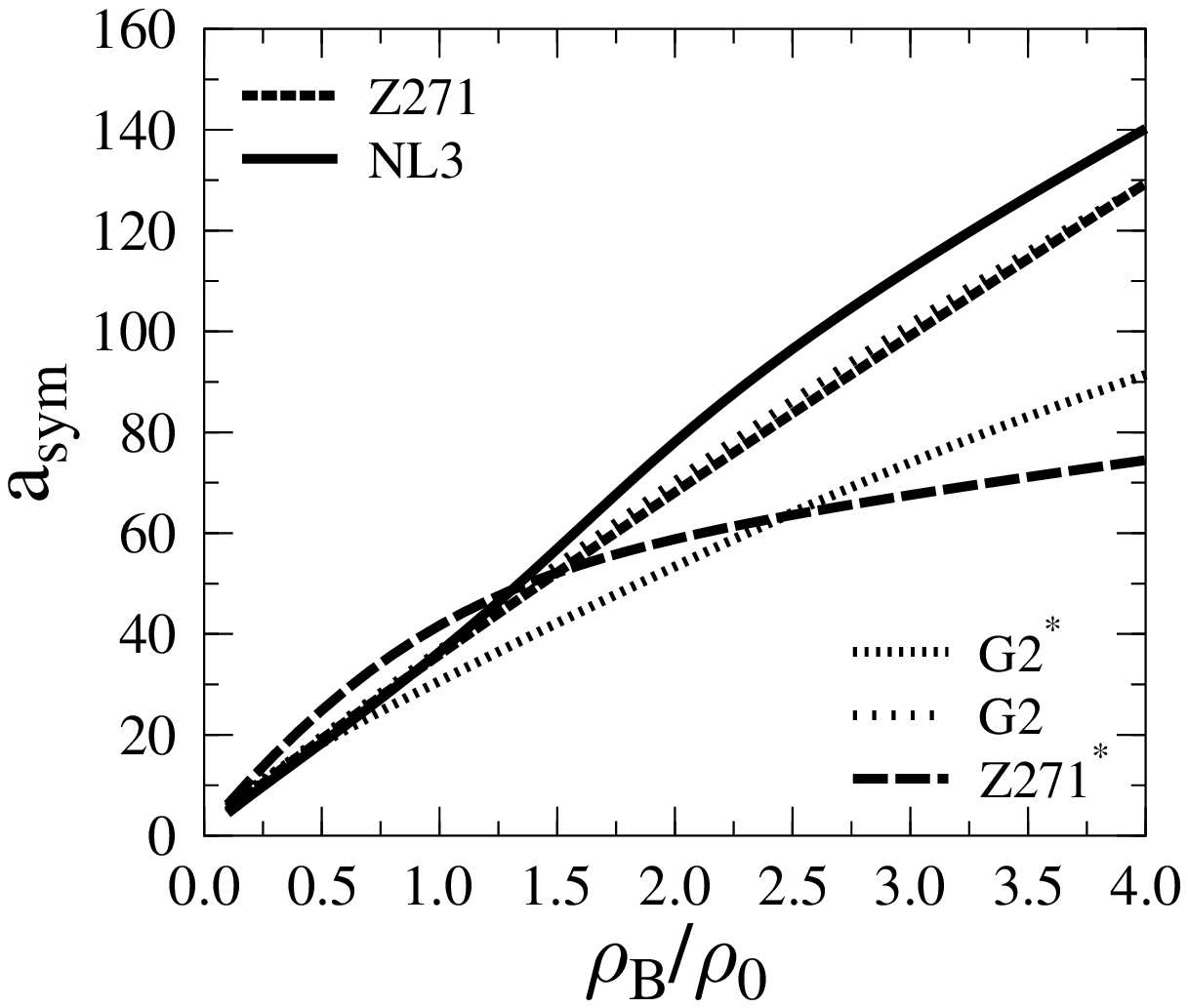,height=7cm}}
\caption{Symmetry energy as a function of the ratio between baryon and nuclear saturation densities for G2, G2*, Z271, Z271* and NL3 parameter sets.}
\label{tab:eden4}
\end{figure*}

In Fig.~\ref{tab:eden2} we show the $\Delta E_2/A$ and $\Delta
E_1/A$ as a function of the ratio between baryon and nuclear
saturation densities for the matter with  and without neutrino
trapping  where G2, G2*, Z271, Z271* and NL3 parameter sets are
used. The difference between the two types of matters appears
mainly at densities lower than 2$\rho_0$, where $\Delta E_2/A$ is
smaller than  $\Delta E_1/A$ for neutrinoless matter. The opposite
situation happens for the neutrino trapping case. In the case of
neutrinoless matter and parameter sets with a stiff $a_{\rm sym}$,
$\Delta E_2/A$ is larger than $\Delta E_1/A$ for the density
larger than 2$\rho_0$,  but for the case of soft $a_{\rm sym}$
this condition is reached only after the density becomes
relatively high.  Figure~\ref{tab:eden3} shows the characteristic
of $\alpha^2$ for both matters as the baryon density increases.
Here, the  dependence of $\alpha^2$  on parameter sets is obvious
for matter without neutrino trapping, while the opposite situation
happens for matter with neutrino trapping. The symmetry energy is
shown in Fig.~\ref{tab:eden4}, where we can clearly see its
dependence on the parameter sets. The dependency of  $\Delta
E_2/A$, $\Delta E_1/A$ and  $\alpha^2$ on $a_{\rm sym}$ can be
investigated by comparing Figs.~\ref{tab:eden2} and
~\ref{tab:eden3} with Fig.~\ref{tab:eden4}. It is obvious that in
the case of matter with neutrino trapping,  correlations between
$\Delta E_2/A$, $\Delta E_1/A$ and  $\alpha^2$ with  $a_{\rm sym}$
at densities lower than 2$\rho_0$ can not be observed, but for the
neutrinoless case the opposite situation appears. For densities
larger than  2$\rho_0$, correlations between $\Delta E_2/A$,
$\Delta E_1/A$, as well as $\alpha^2$ and  $a_{\rm sym}$ for
matter with neutrino trapping is weaker than those without
neutrino trapping.

Therefore, it is clear that the results displayed in
Fig.~\ref{eos1} are caused by one reason: matter without neutrino
trapping has a weak correlation with $a_{\rm sym}$  and especially
at densities lower than  2$\rho_0$, the effect is more pronounced.
This effect is mainly due to the behavior of the $\Delta E_2/A$
contribution. For densities lower than  2$\rho_0$, the asymmetry
between protons and neutrons, which is smaller compared to the
case of neutrinoless matter, also induces a correlation between
$\Delta E_1/A$ and $a_{\rm sym}$, whereas the role of $\Delta
E_1/A$ is suppressed.

In the upper right panel of Fig.~\ref{eos1}, $Y_{\nu_e}$ for different parameter sets are shown. It is interesting to note that  NL3,  Z271 and G2 parameter sets have a similar  $Y_{\nu_e}$ trend but  $Y_{\nu_e}$ of  G2* and  Z271* behaves differently. By comparing this figure with Fig.~\ref{tab:eden4} we can conclude that the number of trapped neutrino correlates with the $a_{\rm sym}$ behavior of the models. This is due to the neutrino fraction which self consistently depends on the proton fraction ($Y_{p}$)  through the neutrality and beta equilibrium conditions, while $Y_{p}$ has a correlation with  $a_{\rm sym}$, although the effect is quite small for this kind of matter.

\begin{figure*}
\centering
\mbox{\epsfig{file=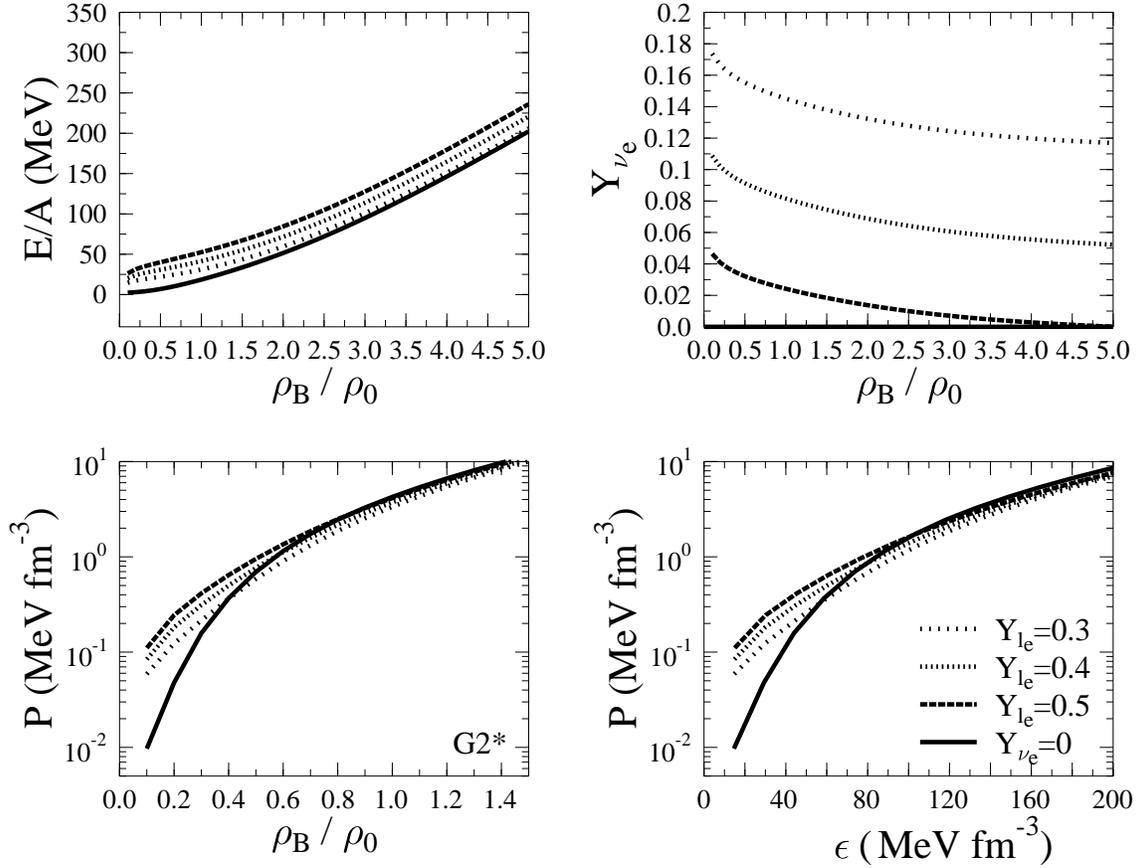,height=12.0cm}}
\caption{Same as Fig.~\ref{eos1}, but here the $Y_{l_e}$ is varied and the G2$^*$ parameter set is used. \label{eos2}}
\end{figure*}

\begin{figure*}
\centering
 \mbox{\epsfig{file=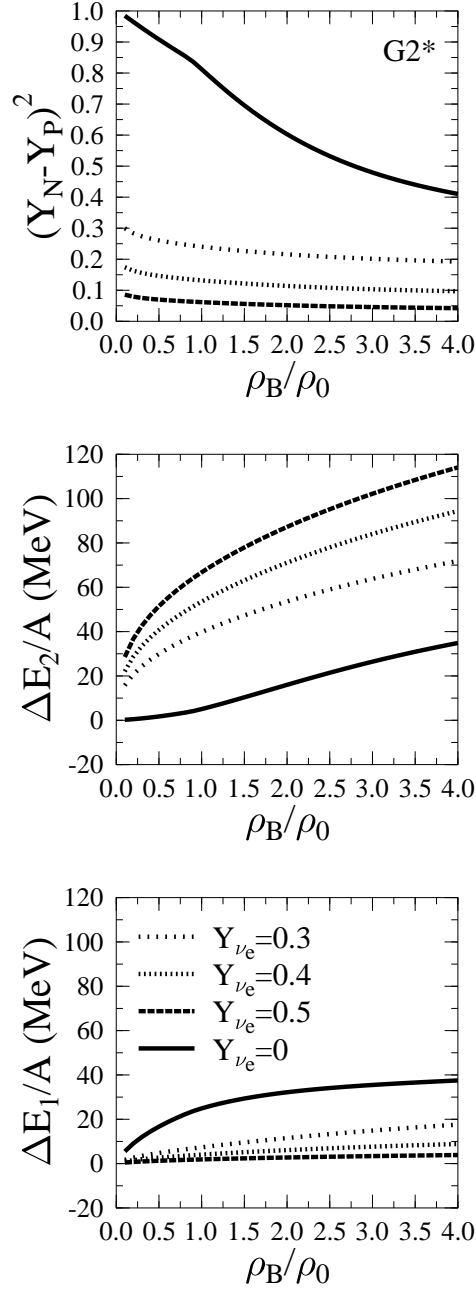,height=18.0cm}}
\caption{$\Delta E_2/A$, $\Delta E_1/A$ and $\alpha$ as a function of the ratio between baryon density and nuclear saturation density for matters with neutrino trapping (NT) obtained by using the G2* parameter set but with varied $Y_{l_e}$.}
\label{tab:eden5}
\end{figure*}

To study the role of the neutrino number in the $E/A$ and  EOS
using G2$^*$ parameter set, we show in the upper left panel of
Fig.~\ref{eos2} the variation of $E/A$ for some values of
$Y_{l_e}$ and in the lower panels the variation of their EOS. It
is found that the larger the number of neutrinos in matter, the
smaller the value of $E/A$. At low density, the EOS of matter with
neutrino trapping is stiffer than  that without neutrino trapping.
According to Ref.~\cite{chiapparini} the reason is that the
neutrino in matter shifts the threshold of muon production toward
higher densities. However, for matter with neutrino trapping,  the
larger the number of neutrinos,  the softer its EOS. At high
densities the situation is opposite. Also we can see this from
another point of view, i.e., by comparing the slope of dominant
contributions of each case (bottom and center panels of
Fig.~\ref{tab:eden5}). At densities less than $\rho_0$, the slope
of  $\Delta E_1/A$ for neutrinoless matter is smaller than that of
$\Delta E_2/A$ for matter with neutrino trapping. In the latter,
also in the same density range, even though it does not clearly
visible, the slope becomes larger as $Y_{l_e}$ becomes larger. The
situation is reversed once the densities become  higher than
$\rho_0$.

\begin{figure*}
\centering
 \mbox{\epsfig{file=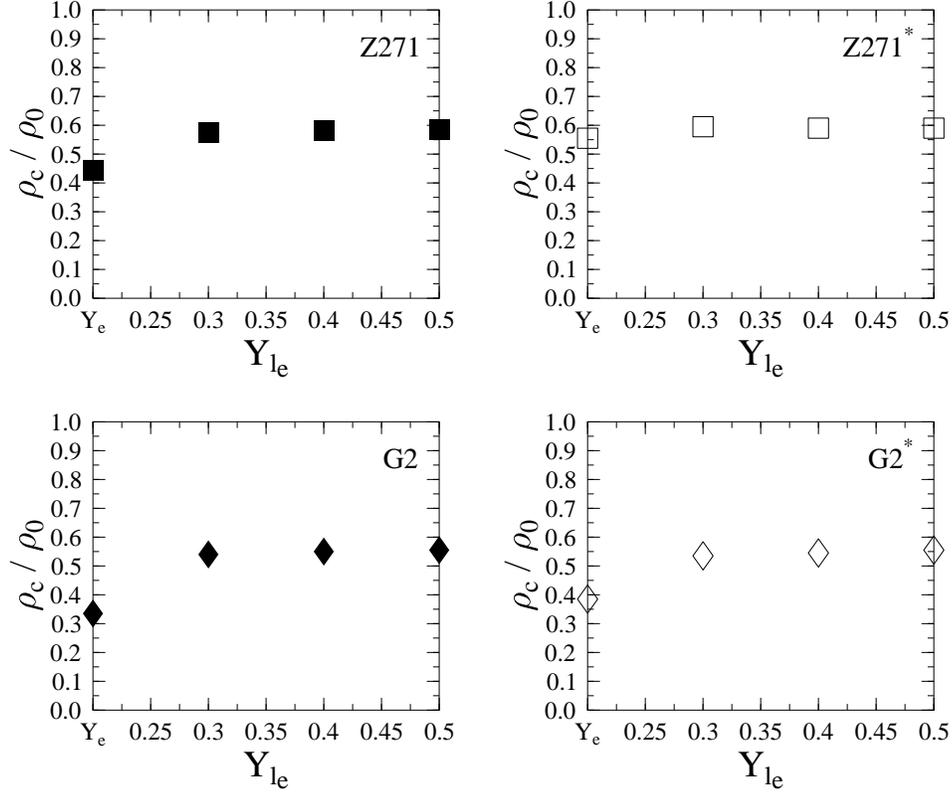,height=11.0cm}}
\caption{Critical densities for the  Z271, Z271*, G2 and G2* parameter sets as a function of the neutrino fraction in matter.\label{cdens1}}
\end{figure*}
\begin{figure*}
\centering
 \mbox{\epsfig{file=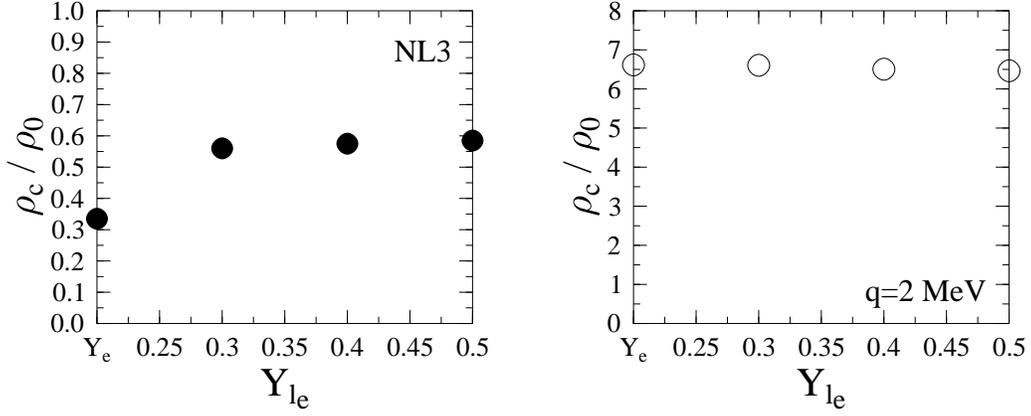,height=6.0cm}}
\caption{Same as in Fig.~\ref{cdens1}, but for the NL3 parameter set.\label{cdens2}}
\end{figure*}

\begin{figure*}
\centering
 \mbox{\epsfig{file=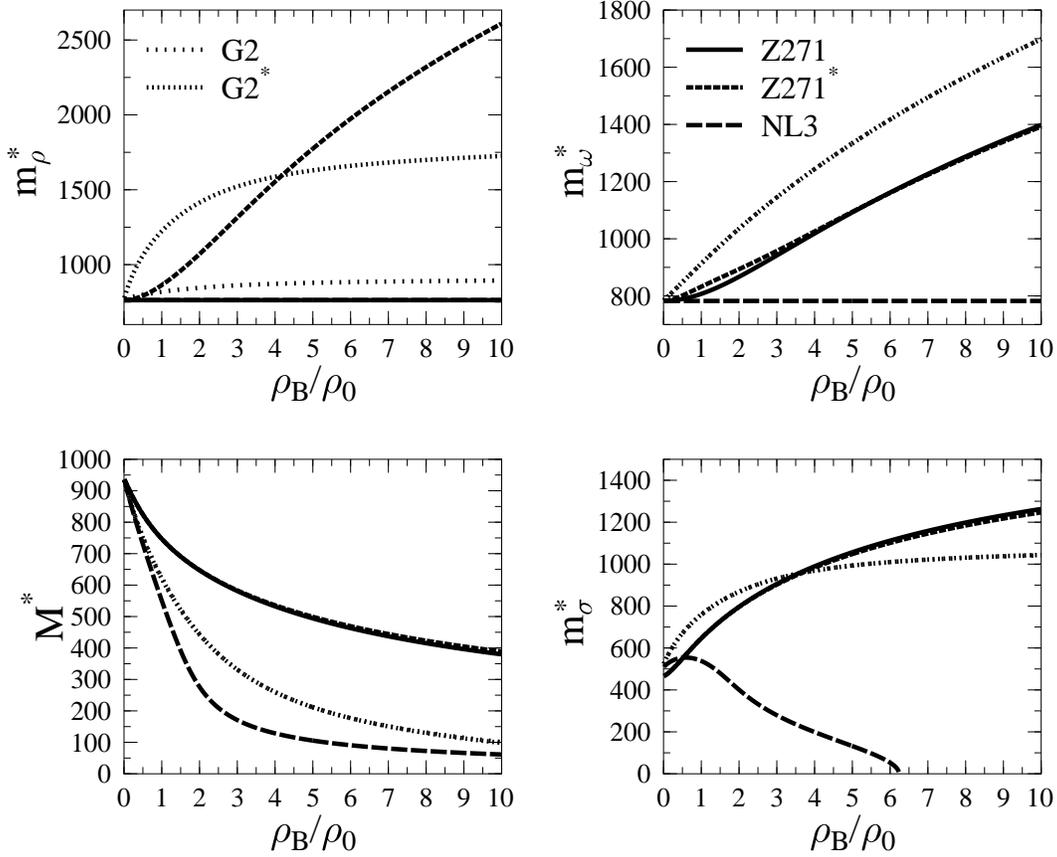,height=12.0cm}}
\caption{Effective masses of nucleon, $\sigma$, $\omega$ and $\rho$ mesons as a function of the baryon density. Note that several lines coincide. \label{masses}}
\end{figure*}

The critical densities as a function of the neutrino fraction in matter are shown in Fig.~\ref{cdens1} for the  Z271 (upper left panel), Z271* (upper right panel), G2 (lower left panel), G2* (lower right panel) and  in the left panel of Fig.~\ref{cdens2} for the NL3 parameter set. Without neutrino trapping, it is obtained that
$\rho_c^{\rm G2}~=~0.052~{\rm fm}^{-3} $, $\rho_c^{\rm G2^*}~=~0.060~{\rm fm}^{-3} $,
$\rho_c^{\rm Z271}~=~0.066~{\rm fm}^{-3} $,  $\rho_c^{\rm Z271^*}~=~0.082~{\rm fm}^{-3} $ and
$\rho_c^{\rm  NL3}~=~0.049~{\rm fm}^{-3} $. This indicates that the critical density in the Horowitz-Piekarewicz model is larger than that of the ERMF and the NL3 parameter sets, while  the standard RMF model represented by the NL3 parameter set yields the smallest value. The effects of the isovector-vector adjustment on $\rho_c$ can be seen by comparing the panels for Z271 with Z271*, or G2 with  G2*. In the Horowitz-Piekarewicz model, the variation of the $\rho_c$ values is larger than in the ERMF one.

On the other hand, with neutrino trapping we obtain,
 $\rho_c^{\rm G2}~=~(0.084-0.086)~{\rm fm}^{-3} $,
$\rho_c^{\rm G2^*}~=~(0.083-0.086)~{\rm fm}^{-3} $,
$\rho_c^{\rm Z271}~=~(0.085-0.086)~{\rm fm}^{-3} $,
$\rho_c^{\rm Z271^*}~=~(0.087-0.088)~{\rm fm}^{-3} $ and
$\rho_c^{\rm NL3}~=~(0.081-0.084)~{\rm fm}^{-3} $.
This indicates that for this case the value of $\rho_c$ does not  sensitively depend on the model and on the variation of the number of trapped  neutrinos in matter. In general, their values are larger than those obtained in  the case without neutrino trapping.

Equation~(\ref{eq:det}) implies that $\rho_c$ is determined by matter composition as well as  by the effective masses of mesons and nucleons used. We know from previous results that protons and neutrons are the dominant constituents in matter composition and its behavior can be observed from the corresponding asymmetry ($\alpha$) which has been discussed and shown in Figs.~\ref{tab:eden3} and~\ref{eos2}. On the other hand, the effective masses depend also on the used parameter sets (see Fig.~\ref{masses}).  The above findings and the facts that for each parameter set $\rho_c^{\rm with ~NT} < \rho_c^{\rm without ~NT}$, as well as the indication that in matter with neutrino trapping and constant $Y_{l_e}$ the critical density does not vary significantly for all parameter sets used, can be understood as the evidence of the dominant role played by the proton-neutron asymmetry in matter in determining the value of $\rho_c$. Furthermore, the reason that the strong dependence of  $\rho_c$ on the parameter set used in the case of matter without neutrino trapping is that the protons-neutrons asymmetry of this kind of matter has a strong correlation with the $a_{\rm sym}$.

From Figs.~\ref{eos2},~\ref{cdens1} and the top panel of Fig.~\ref{tab:eden5} we can clearly see that the  dependence of  $\rho_c$ on the number of trapped neutrinos in matter does not significantly change with respect to the variation of $Y_{l_e}$ and the value of $\alpha^2$. This result emphasizes our previous finding that there is a strong relation between $\rho_c$ and the $a_{\rm sym}$.

To show the crucial role of protons in shifting the values of
$\rho_c$ to relatively low densities, let us study the
contribution of each constituent. First, the electrons, protons,
and muons contributions are switched off. Then, by fixing $q$  at
30 MeV we search  for $\rho_c$ for all parameter sets, both for
trapped and untrapped neutrino cases. For all cases, we found
more or less similar results, i.e., $\rho_c/\rho_0 ~\approx~ 1.7$.
This value comes merely from the neutron contribution. A small
difference in the neutron fraction ($Y_n$) for both cases has
a negligible effect on the $\rho_c$ prediction. Secondly, now in
addition to the first condition, the proton contribution is
switched on. Now we obtain quite different values of $\rho_c$, for
example, in the case of G2* with  $Y_{l_e}=0.3$ we obtain
$\rho_c/\rho_0=0.580$, and for the case  without neutrino trapping
we obtain $\rho_c/\rho_0=0.425$. Thus, it is clear that the
protons shift the  $\rho_c$ values to the region with relatively
low densities. Next, the electron contribution is  switched on.
Then  we observed that $\rho_c/\rho_0$ is shifted further, i.e.,
$\rho_c/\rho_0=0.430$ for $Y_{l_e}=0.3$ and $\rho_c/\rho_0=0.350$
for matter without neutrino trapping. If the muons contribution is
switched on, the result does not change. It happens not only in
matter with neutrino trapping, but also in matter without neutrino
trapping, respectively. However, these results show that the proton
and electron contributions decrease the value of $\rho_c$ in
different ways for both cases and, therefore, reveal the crucial
role of matter composition  in determining the value of $\rho_c$.

In contrast to the Horowitz-Piekarewicz and ERMF models, the NL3
parameter set yields an additional instability at relatively high
densities. This is shown in the right panel of Fig.~\ref{cdens2}
for the $q$ = 2 MeV case, i.e., $\rho_c/\rho_0 ~\approx~ 6.5$. Even
if we use $q$ close to zero, this instability does not disappear
and the $\rho_c/\rho_0$ value stays similar as in the $q$ = 2 MeV
case. Therefore, this instability is not due to the small-amplitude
density fluctuations. In fact, this instability appears because
the effective $\sigma$ mass of the NL3 parameter set is zero at
that density (as shown in the lower right panel of
Fig.~\ref{masses}). As a consequence,  the corresponding $\sigma$
propagator changes the sign at this point. This fact clearly shows
that additional nonlinear terms to the $\sigma$ nonlinearities of the 
standard RMF models like in the Horowitz-Piekarewicz and ERMF ones 
can prevent such kind of instability to appear.

\section{Conclusion}
\label{sec_conclu}
The effects of the different treatments in the isovector-vector
sector of RMF models on the properties of matter with neutrino
trapping have been studied. The effects are less significant
compared to those without neutrino trapping. Different dependences
of the EOS and $B/A$ of both matters on $a_{\rm sym}$ are the reason
behind this. 

The effects of the variation of the  neutrino
fraction in matter on the EOS and $B/A$ have been also discussed.

The longitudinal dielectric function of the ERMF model for matter
consisting of neutrons, protons, electrons, muons, and neutrinos
has been derived. The result is used to study the dynamical
instability of uniform matters at low densities. The behavior of
the predicted $\rho_c$ in matter with and without neutrino
trapping has been investigated. It is found that different
treatments in the isovector-vector sector of RMF models yield more
substantial effects in matter without  neutrino trapping rather
than in matter with neutrino trapping. Moreover, for  matter with
neutrino trapping, the value of $\rho_c$ does not significantly
change with the variation of the models as well as  with the
variation of the neutrino fraction in matter. In this case, the
value of $\rho_c$ is larger for matter with neutrino trapping.
These are due to the interplay between the major role of matter
composition and the role of the effective masses of mesons and
nucleons. It is also found that the additional nonlinear terms of
Horowitz-Piekarewicz and ERMF models prevent another instability
at relatively high densities  to appear. This can be traced
back to the effective $\sigma$ mass which goes to zero when the
density approaches $6.5~\rho_0$ .
\section*{ACKNOWLEDGMENT}
We are indebted to Marek Nowakowski for useful suggestions and
proofreading this manuscript. We also acknowledge the support from
the Hibah Pascasarjana grant as well as from the Faculty of
Mathematics and Sciences, University of Indonesia.

\appendix
\section{Meson Propagators}
\label{sec_mp}
In this Appendix, simplified Dyson equations for meson propagators (only zero component of each propagator is considered) will be given. The covariant form for $\sigma$ and $\omega$ couplings  is given in Ref.~\cite{Celenza}. If we define the $\sigma$, $\omega$ and $\rho$ free propagators as
\be
G_\sigma = \frac{i}{q_\mu^2-m_\sigma^2}~ ~ ~ ~ ~ G_\omega^{00} = \frac{-i}{q_\mu^2-m_\omega^2}~ ~ ~ ~ ~ G_\rho^{00} = \frac{-i}{q_\mu^2-m_\rho^2},
\ee
the Dyson equation for the $\sigma$ propagator in the absence of coupling to $\omega$ and $\rho$ fields is obtained by considering the sum of ring diagrams, i.e.,
\be
\bar{G}_\sigma = G_\sigma - i  G_\sigma \Pi_{\sigma \sigma} \bar{G}_\sigma,
\ee
from which we can obtain
\be
\bar{G}_\sigma = \frac {G_\sigma} {1+ i  \Pi_{\sigma \sigma}G_\sigma }=\frac{i}{q_\mu^2-(m^*_\sigma)^2}.
\ee
Similarly, the Dyson equations for the zero components of $\omega$ and $\rho$ meson propagators are
\be
\bar{G}_\omega^{00} = \frac {G_\omega^{00}} {1- i  \Pi_{\omega \omega}^{00} G_\omega^{00} }=\frac{-i}{q_\mu^2-(m^*_\omega)^2},
\ee
and
\be
\bar{G}_\rho^{00} = \frac {G_\rho^{00}} {1- i  \Pi_{\rho \rho}^{00} G_\rho^{00} }=\frac{-i}{q_\mu^2-(m^*_\rho)^2},
\ee
where $(m^*_\sigma)^2$= $(m_\sigma)^2$ +$\Pi_{\sigma \sigma}$, $(m^*_\omega)^2$= $(m_\omega)^2$ +$\Pi_{\omega \omega}^{00}$, and $(m^*_\rho)^2$= $(m_\rho)^2$ +$\Pi_{\rho \rho}^{00}$.

In the ERMF model, each meson is coupled to other mesons through the nonlinear terms. This fact further complicates the form of the full propagators.
The Dyson equation for the $\sigma$ propagator including the possible mixing terms becomes
\bea
\tilde{G}_\sigma=\bar{G}_\sigma-\bar{G}_\sigma \Pi_{\sigma \omega}^0 \bar{G}_\omega^{00}\Pi_{\sigma \omega}^0 \tilde{G}_\sigma-\bar{G}_\sigma \Pi_{\sigma \rho}^0 \bar{G}_\rho^{00}\Pi_{\sigma \rho}^0 \tilde{G}_\sigma ,
\eea
which can be written as
\be
\tilde{G}_\sigma=\frac{\bar{G}_\sigma}{1+\bar{G}_\sigma(\Pi_{\sigma \omega}^0 \bar{G}_\omega^{00}\Pi_{\sigma \omega}^0+\Pi_{\sigma \rho}^0 \bar{G}_\rho^{00}\Pi_{\sigma \rho}^0)}.
\label{eq:fullprop}
\ee
Equation~(\ref{eq:fullprop}) can be simplified into
\bea
\tilde{G}_\sigma=\frac{i (q_\mu^2-m_\omega^{*~2})(q_\mu^2-m_\rho^{*~2})}{(q_\mu^2-m_\omega^{*~2})(q_\mu^2-m_\rho^{*~2})(q_\mu^2-m_\sigma^{*~2})+(\Pi_{\sigma \omega}^0)^2(q_\mu^2-m_\rho^{*~2})+(\Pi_{\sigma \rho}^0)^2 (q_\mu^2-m_\omega^{*~2})}.
\label{eq:prop1}
\eea
Similarly, the zero component of $\omega$ and $\rho$ meson propagators can be written as
\bea
\tilde{G}_\omega^{00}=\frac{-i (q_\mu^2-m_\sigma^{*~2})(q_\mu^2-m_\rho^{*~2})}{(q_\mu^2-m_\omega^{*~2})(q_\mu^2-m_\rho^{*~2})(q_\mu^2-m_\sigma^{*~2})+(\Pi_{\sigma \omega}^0)^2(q_\mu^2-m_\rho^{*~2})-(\Pi_{\omega \rho}^{00})^2 (q_\mu^2-m_\sigma^{*~2})},
\label{eq:prop2}
\eea
and
\bea
\tilde{G}_\rho^{00}=\frac{-i (q_\mu^2-m_\sigma^{*~2})(q_\mu^2-m_\omega^{*~2})}{(q_\mu^2-m_\omega^{*~2})(q_\mu^2-m_\rho^{*~2})(q_\mu^2-m_\sigma^{*~2})+(\Pi_{\sigma \rho}^0)^2(q_\mu^2-m_\omega^{*~2})-(\Pi_{\omega \rho}^{00})^2 (q_\mu^2-m_\sigma^{*~2})}.~~
\label{eq:prop3}
\eea
We can also define a propagator $\tilde{G}_{\omega \sigma}^{0}$ which contains the sum of all diagrams that transform $\omega$ into  $\sigma$, i.e.,
\bea
\tilde{G}_{\omega \sigma}^{0}&=& - i \bar{G}_{\omega}^{00} \Pi_{ \omega\sigma}^0\bar{G}_\sigma + i \bar{G}_{\omega}^{00} \Pi_{ \omega\sigma}^0\bar{G}_\sigma  \Pi_{ \sigma \omega}^0\bar{G}_{\omega}^{00}\Pi_{ \omega\sigma}^0\bar{G}_\sigma+ \cdot \cdot \cdot\nonumber\\ &+&i \bar{G}_{\omega}^{00} \Pi_{ \omega\sigma}^0\bar{G}_\sigma  \Pi_{ \sigma \rho}^0\bar{G}_{\rho}^{00}\Pi_{ \rho\sigma}^0\bar{G}_\sigma+ \cdot \cdot \cdot +i \bar{G}_{\omega}^{00} \Pi_{ \omega\rho}^0\bar{G}_\rho^{00}  \Pi_{ \rho \omega}^0\bar{G}_{\omega}^{00}\Pi_{ \omega\sigma}^0\bar{G}_\sigma+ \cdot \cdot \cdot\nonumber\\ &+&\cdot \cdot \cdot.
\eea
This propagator may be summed up to produce
\bea
\tilde{G}_{\omega \sigma}^{0}=\frac{-i \bar{G}_{\omega}^{00} \Pi_{ \omega\sigma}^0\bar{G}_\sigma}{1+\bar{G}_\sigma  \Pi_{ \sigma \omega}^0\bar{G}_{\omega}^{00}\Pi_{ \omega\sigma}^0+\bar{G}_\sigma  \Pi_{ \sigma \rho}^0\bar{G}_{\rho}^{00}\Pi_{ \rho\sigma}^0+\bar{G}_\omega^{00}  \Pi_{ \omega \rho}^{00}\bar{G}_{\rho}^{00}\Pi_{\rho \omega}^{00}},
\eea
which can be simplified to
\bea
\tilde{G}_{\omega \sigma}^{0}=\frac{-i \Pi_{\omega \sigma}^{0}(q_\mu^2-m_\rho^{*~2})}{H(q,q_0)}.
\label{eq:prop4}
\eea
Similarly, we can also obtain  the sum of all diagrams that transform  $\rho$ into  $\sigma$ as
\bea
\tilde{G}_{\rho \sigma}^{0}=\frac{-i \Pi_{\rho \sigma}^{0}(q_\mu^2-m_\omega^{*~2})}{H(q,q_0)},
\label{eq:prop5}
\eea
and  that transform $\rho$ into  $\omega$ as
\bea
\tilde{G}_{\rho \omega}^{00}=\frac{i \Pi_{\rho \omega}^{00}(q_\mu^2-m_\sigma^{*~2})}{H(q,q_0)},
\label{eq:prop6}
\eea
where
\bea
H(q,q_0)&=&(q_\mu^2-m_\omega^{*~2})(q_\mu^2-m_\rho^{*~2})(q_\mu^2-m_\sigma^{*~2})+(\Pi_{\sigma \omega}^0)^2(q_\mu^2-m_\rho^{*~2})+(\Pi_{\sigma \rho}^0)^2 (q_\mu^2-m_\omega^{*~2})\nonumber\\&-&(\Pi_{\omega \rho}^{00})^2 (q_\mu^2-m_\sigma^{*~2}).
\eea
Equations~(\ref{eq:d1})~-~(\ref{eq:d6}) are special cases  of  Eqs.~(\ref{eq:prop1}~-~\ref{eq:prop3},~\ref{eq:prop4},~\ref{eq:prop5},~\ref{eq:prop6}), i.e. by taking the limit of $q_0 \rightarrow$ 0 and inserting the proper mesons coupling constants in the latter.
\section{Explicit Form of the Longitudinal Dielectric Function}
\label{sec_det}
In this Appendix, the explicit form of the longitudinal dielectric function $\epsilon_L$ = $\[1-D_L(q) \Pi_L(q,q_0=0)\]$ is provided.
If we define the matrix $\epsilon_L$ as
\bea
\epsilon_L = \left( \begin{array} {ccccc}
A_1 & B_1 & D_1&E_1&F_1\\
A_2 & B_2 & D_2&E_2&F_2\\
A_4 & B_4 & D_4&E_4&F_4\\
A_5 & B_5 & D_5&E_5&F_5\\
A_6 & B_6 & D_6&E_6&F_6\\
\end{array}\right),
\label{eq:detex1}
\eea
then the contents of each component of the matrix in Eq.~(\ref{eq:detex1}) are
\bea
\begin{array} {lll}
A_1~=~1-d_g \Pi_{00}^e , &A_2~=~-d_g \Pi_{00}^e , &A_4~=~0,\\
B_1~=~ -d_g \Pi_{00}^{\mu} , &B_2~=~1-d_g \Pi_{00}^{\mu} , &B_4~=~0,\\
D_1~=~ d_g \Pi_{m}^p , &D_2~=~ d_g \Pi_{m}^p , &D_4~=~ 1+d_s \Pi_{s}-d_{sv\rho}^+ \Pi_{m}^p-d_{sv\rho}^- \Pi_{m}^n,\\
E_1~=~ d_g \Pi_{00}^p , &E_2~=~ d_g \Pi_{00}^p , &E_4~=~ d_s \Pi_{m}^p-d_{sv\rho}^+\Pi_{00}^p,\\
F_1~=~ 0 , &F_2~=~ 0 , &F_4~=~ d_s \Pi_{m}^n-d_{sv\rho}^-\Pi_{00}^n,
\end{array}
\eea
and
\bea
\begin{array} {lll}
A_5= d_g \Pi_{00}^e,~~~~~&A_6=0,\\
B_5= d_g \Pi_{00}^{\mu},~~~~~&B_6=0,\\
D_5= -d_{sv\rho}^+\Pi_{s}-d_{33}\Pi_{m}^p-d_{v\rho}^-\Pi_{m}^n, ~~~~~&D_6=-d_{sv\rho}^-\Pi_{s}-d_{v\rho}^-\Pi_{m}^p-d_{44}\Pi_{m}^n,\\
E_5=1-d_{sv\rho}^+\Pi_{m}^p-d_{33}\Pi_{00}^p,~~~~~&E_6= -d_{sv\rho}^-\Pi_{m}^p-d_{v\rho}^-\Pi_{00}^p,\\
F_5= -d_{sv\rho}^+\Pi_{m}^n-d_{v\rho}^-\Pi_{00}^n,~~~~~&F_6=1-d_{sv\rho}^-\Pi_{m}^n-d_{44}\Pi_{00}^n .
\end{array}
\eea


\end{document}